# SEMI-PHENOMENOLOGICAL CLASSIFICATION MODELS OF THE GENETIC CODE(S) USING $q$-DEFORMED NUMBERS

Tidjani Négadi

Department of Physics, Faculty of Science, University of Oran, 31100, Es-Sénia, Oran, Algeria, Email address: tnegadi@gmail.com

**Abstract**: The mathematical concept of *q*-deformations, in particular the one of *q*-numbers, is used to study the genetic code(s). After considering two kinds of *q*-numbers, for comparison, a phenomenological classification scheme of the genetic code together with its numerous minor variants is, first, established. Next, numbers describing the presence of additional amino acids, such as Selenocysteine or/and Pyrrolysine, are also produced. Finally, a minimal number of amino acids, which could fit the small number of them which are thought to have been involved, at the origin of life on Earth, is found. All together, these results constitute our final semi-phenomenological model.

**Keywords**: *q*-deformations, *q*-numbers, genetic code(s), Selenocysteine, Pyrrolysine, primordial amino acids

## 1. INTRODUCTION

This work is devoted to the application of what are today known as *q*-deformations to the study of the genetic code(s). The concept of *q*-deformations, or "quantum" deformations, is well known in mathematical and theoretical physics and has been widely applied to the study of various physical systems, harmonic oscillator (Macfarlane, 1989), hydrogen atom (Kibler and Négadi, 1991), periodic table of the atomic elements (Négadi and Kibler, 1992), vibrational and rotational spectra of molecules and atomic nuclei, solid-state physics (see, for example Bonatsos and Daskaloyannis (1999), and the references therein), non-linear physics (see, for example,

Batchelor et al., 1990), higher-order interactions in many-body problems (see, for example, Sviratcheva et al., 2004), Yang-Mills Theory (see, for example, Szabo et al., 2013), string theory (see for example Chaichian, 1994), and so on (the interested reader could find many other references by a Google search or by consulting the Cornell Library preprint repository at arXiv.org). The main idea behind q-deformations is to introduce a deformation parameter $q$, which could be real (or complex), in the "classical" mathematical expressions describing the studied system, and look at the effect(s) produced by the variation of the deformation parameter on these expressions. (In some theories more than one deformation parameter are considered, but this does not interest us, here.) In the limit where $q=1$, the "classical" expressions are recovered and, for $q \neq 1$, one expects new insights into the studied systems. The basic mathematical tools of these $q$-deformations are, at the very beginning, objects like $q$-numbers, $q$-deformed elementary functions, $q$-derivative, $q$-integral. This work is therefore dedicated to the application, to a *biological* problem, *i.e.*, the study of the genetic code and its several known existing variants, or non-standard genetic codes, using the all first of these tools, the $q$-numbers. As there exist several definitions of these latter, we consider, in this work, two well known of them, one mainly used in the mathematical literature and the other mainly used in physics research, *for comparison*, and this will lead us to use also the Vertebrate Mitochondrial genetic code, as a starting point, besides the standard genetic code. We will therefore look at the various genetic code variants as *deformations* of, for example, the standard genetic code, as a starting point, this latter being described by its numbers (number of amino acids, number of amino acids in the multiplets, degeneracies). The obtained final result is a mathematical semi-phenomenological classification, *in one equation*, where these genetic code variants, grouped into categories according to their stop-codon numbers, correspond to different $q$-values. Also, this classification could encompass the existence of additional numbers of amino acids, besides the 20 canonical ones and examples for known cases are given. Finally, the above mentioned classification scheme is shown to also possess a certain "*predictive*" capability, namely, the existence of a *minimal number* of amino acids which number, being small, could be favorably compared with old and still persistent today claims, made by several leading authors, concerning the number of amino acids which are thought to have been involved in the primitive genetic code, at the origin of life on Earth. Because of this predictive property, our final model will be termed *semi-phenomenological*. In the second section, we briefly describe the two mentioned kinds of $q$-numbers we will use in this study, for comparison. In the third section, we present the (experimental) numerical description of the standard genetic code and also of the Vertebrate Mitochondrial genetic code which are both involved in this study, which description is also enriched by corresponding mathematical fitting (empirical) *exact* formulas. In a forth section (sub-section 4.1) we present, the case where we start from the standard genetic code and use the two kinds of *q*-numbers. In sub-section 4.2, because of an encountered obstruction (see section 4.1), we do the same but starting from the Vertebrate Mitochondrial genetic code and use also the two kinds of *q*-numbers. In the last section, we summarize our results for our final semi-phenomenological classification model.

## 2. TWO KINDS OF *q*-NUMBERS

We consider here, for comparison in this study, the following two well-known kinds of *q*-numbers, the first, Eq.(I), usually considered in the mathematical literature and the second, Eq.(II), mainly used in physics research:

$$[n]_q = \frac{q^n - 1}{q - 1} \quad \text{(I)}$$

$$[n]_q = \frac{q^n - q^{-n}}{q - q^{-1}} \quad \text{(II)}$$

where *n* is a natural number and $[n]_q \to_{q \to 1} n$, in both cases. Explicitly, for the first kind, (I), we have the (other) general form

$$[n]_q = 1 + q + q^2 + \ldots + q^{n-1} \quad (1)$$

For example $[1]_q = 1$, $[2]_q = 1+q$, $[3]_q = 1+q+q^2$, and so on. For the second kind, (II), the first few terms are given by

$$[1]_q = 1, [2]_q = q + q^{-1}, [3]_q = q^2 + 1 + q^{-2}, \ldots \quad (3)$$

The first one, (I), which is known for a long time (Euler, 1748; Heine, 1878; Jackson, 1904) possesses *no symmetry* while the second, (II), more recently introduced in physics (Macfarlane, 1989), is invariant (*symmetric*) under the substitution $q \to q^{-1}$. This latter fact has some consequences in this work.

## 3. THE STANDARD GENETIC CODE AND ITS VARIANTS

The genetic code is the set of rules by which the genetic information, encoded in the genes, is translated inside the ribosomes into proteins. Specifically, it is a many-to-one mapping between 64 codons and 20 amino acids. In the *standard* genetic code, used by the great majority of living organisms, 61 codons are meaningful, *i.e.*, correspond to amino acids and the remaining 3 codons are termination signals, or stops. Also, as several codons could code for one amino acid, there exist a degeneracy phenomenon and a multiplet structure. In the detail, one has five quartets (five amino acids, each coded by four codons: Proline P, Alanine A, Threonine T, Valine V, Glycine G), nine doublets (nine amino acids, each coded by two codons: Phenylalanine F, Tyrosine Y, Cytosine C, Histidine H, Glutamine Q, Asparagine N, Lysine K, Aspartic Acid D, Glutamic Acid E), three sextets (three amino acids, each coded by six codons: Leucine L, Arginine R, Serine S), one triplet (one amino acid coded by three codons: Isoleucine I) and finally two singlets (two amino acids, each coded by a one codon: Methionine M, Tryptophane W). This numerical structure is shown below where $n_i$ is the number of amino acids coded by $i$ codons (see also Table 1)

$$\text{Quartets: } n_4 = 5, \ i = 4$$
$$\text{Doublets: } n_2 = 9, \ i = 2$$
$$\text{Sextets: } n_6 = 3, \ i = 6 \quad (4)$$
$$\text{Triplet: } n_3 = 1, \ i = 3$$
$$\text{Singlets: } n_1 = 2, \ i = 1$$

The numbers $n_i$, that is the number of amino acids in degeneracy-class $i$ (or *class-number i*), are called *group-numbers*. Interestingly, this (experimental) structure could be supported by *exact* mathematical formulas, from different approaches. We have shown some years ago (Négadi, 2007, 2008, 2009) that the multiplet structure in Eq.(4) could be derived, exactly, from the (unique) number 23!, the order of the permutation group of 23 objects $S_{23}$. There exist also another interesting way to derive, also exactly, this same structure (see Négadi, 2015). It is given by the following sum

$$\sum_{K,k}(2^k + \delta_{K,0}) \times (8 - 2k - 5\delta_{K,1}) \quad (5)$$

where $\delta_{i,j}$ is Kronecker's delta function, equal to 1 for $i=j$ and 0 otherwise. In the above equation, $2^k+\delta_{K,0}$ gives the group-numbers and $8-2k-5\delta_{K,1}$ gives the class-numbers. For $K=0$ and $k=1$, 2 and 3, we get the group-numbers 3, 5 and 9, for the class-numbers 6, 4 and 2, respectively. For $K=1$ and $k=0$ and 1, we have the group-numbers 1 and 2, for the class-numbers 3 and 1, respectively. Expliciting equation (5), we get

$$\sum_{k=1,2,3}(2^k+1)(8-2k) + \sum_{k=0,1}2^k(3-2k) =$$

$$(3\times6+5\times4+9\times2)+(1\times3+2\times1)=(18+20+18)+(3+2)=61 \qquad (6)$$

We have, from Eq.(6), 3 sextets coded by 18 codons, 5 quartets coded by 20 codons, 9 doublets coded by 18 codons, 1 triplet coded by 3 codons and finally 2 singlets each coded by 1 codon. As the total of coding codons is 61, we have therefore 3 stop codons (64-61). As another important example, also considered in this work, let us describe the Vertebrate Mitochondrial genetic code. In this case there are 12 doublets (F, Y, C, H, Q, N, K, D, E, I, W, M), 6 quartets (P, A, T, V, G, R) and 2 sextets (L, S) with a total of 60 codons to code the amino acids and, consequently, 4 stop codons left (64-60). Here, also, we could write a mathematical formula to describe the above degeneracy pattern (see Table 1). It is given by the following sum

$$\sum_{k=0,1,2}\left(2^{(k+1)}+2k\right)(6-2k) =$$

$$(2\times6+6\times4+12\times2)=(12+24+24)=60 \qquad (7)$$

The first term, $2^{(k+1)}+2k$, gives the number of amino acids (2, 6 and 12) for the degeneracy-class numbers 6, 4 and 2, respectively and given by the second term $6-2k$. In this case, we have, as mentioned above, 4 stop codons. Now, it is well known that, besides the standard genetic code, there exist several (minor) *variants* genetic codes (Elzanowski and Ostell, 2013), also called non-standard genetic codes. They all involve the *same* set of 20 amino acids but use the set of 64 codons slightly differently. Their degeneracy patterns are shown below, in Table 1. In the first line of the Table, the number of codons used for coding an amino acid, are indicated (1 for singlets, 2 for doublets, 3 for triplets, and so on). In the rest of the Table and for each variant, the number of amino acids, for each degeneracy-class, are indicated (see Eqs. (6) and (7) for the cases of the Standard Genetic Code and the Vertebrate Mitochondrial Genetic Code.

| Number of codons | 1 | 2 | 3 | 4 | 5 | 6 | 7 | 8 | # stops |
|---|---|---|---|---|---|---|---|---|---|
| The Vertebrate Mitochondrial Code | | 12 | | 6 | | 2 | | | 4 |
| The Thraustochytrium Mitochondrial Code | 2 | 9 | 1 | 5 | 1 | 2 | | | 4 |
| The Standard Code | 2 | 9 | 1 | 5 | | 3 | | | 3 |
| The Bacterial, Archeal and plant Plastide Code | 2 | 9 | 1 | 5 | | 3 | | | 3 |
| The Alternative Yeast Nuclear Code | 2 | 9 | 1 | 5 | 1 | 1 | 1 | | 3 |
| The Scenedesmus obliqus Mitochondrial Code | 2 | 9 | 1 | 5 | 1 | 1 | 1 | | 3 |
| The Yeast Mitochondrial Code | | 13 | | 4 | | 2 | | | 2 |
| The Mold, Protozoan, and Coelenterate Code and the Mycoplasma/Spiroplasma Code | 1 | 10 | 1 | 5 | | 3 | | | 2 |
| The Invertebrate Mitochondrial Code | | 12 | | 6 | | 1 | | 1 | 2 |
| The Echinoderm and Flatworm Mitochondrial Code | 2 | 8 | 2 | 6 | | 1 | | 1 | 2 |
| The Euploid Nuclear Code | 2 | 8 | 2 | 5 | | 3 | | | 2 |
| The Ascidian Mitochondrial Code | | 12 | | 5 | | 3 | | | 2 |
| The Chlorophycean Mitochondrial Code | 2 | 9 | 1 | 5 | | 2 | 1 | | 2 |
| The Trematode Mitochondrial Code | 1 | 10 | 1 | 6 | | 1 | | 1 | 2 |
| The Pterobranchia Mitochondrila Code | 2 | 10 | 1 | 6 | | 1 | 1 | | 2 |
| The Candidate Division SR1 and Gracilibacteria Code | 2 | 9 | 1 | 4 | 1 | 3 | | | 2 |
| The Ciliate, Dasycladacean and Hexamita Nuclear Code | 2 | 8 | 1 | 6 | | 3 | | | 1 |
| The Alternative Mitochondrial Code | 2 | 7 | 3 | 6 | | 1 | | 1 | 1 |

**Table 1**. The standard and non-standard genetic codes with their degeneracy patterns

In the above table, we see that there are *four categories*, $C_i$, of variant genetic codes, according to the number of stop codons: i=4, 3, 2 and 1 (last column). Below, we shall use this categorization in our classifications, as the number of stop codons is a property *shared* by all the variant genetic codes in a given category (see the remark below, after Eq.(9)). Now, the equation (6) above, for the standard genetic code, could be rearranged as follows

$$[\sum_{k=1,2,3}(2^k+1) + \sum_{k=0,1}2^k] + [\sum_{k=1,2,3}(2^k+1)(8-2k-1) + \sum_{k=0,1}2^k(3-2k-1)] = \quad (8)$$

$$[(3+5+9)+(1+2)] + [(15+15+9)+(2+0)] = 20 + 41 = 61$$

The two sums in the first bracket give the total number of amino acids 3 sextets, 5 quartets, 9 doublets, 1 triplet and 2 singlets. The two sums in the second bracket give the respective degeneracies. The way we count the degeneracies is as follows. Take, for example, the three sextets coded by 18 codons. For *each* sextet, we have six coding codons. Pick *any one* of them and call it "*amino acid*". It remains five other codons which are termed *degenerate*. For the three sextets we have therefore 3 "*amino acid*" and 3×(6-1)=15 *degenerate* codons. Now, we know that there are 64 codons in all so that, by subtracting the second member of Eq.(8) from 64, we have

$$64 - (20 + 41) = 3 \qquad (9)$$

This is the number of codons not used for coding the amino acids, *i.e.*, the number of stop codons, or stops. Note that this categorization in terms of stop-codons is equivalent to the one in terms of the total number of degenerate codons as it could be seen, from the table, that all the variant genetic codes in a given category, although having different group-numbers and class-numbers, they have the *same* total degeneracy number. For example, in category **C₄**, on has 12×1+6×3+2×5=2×0+9×1+1×2+5×3+1×4+2×5=40 (see the remark, just above Eq.(9), about how we count the degeneracies). Proceeding the same way for the Vertebrate Mitochondrial genetic code in Eq.(7), we get

$$\sum_{k=0,1,2} \left(2^{(k+1)} + 2k\right) + \sum_{k=0,1,2} \left(2^{(k+1)} + 2k\right)(6 - 2k - 1) =$$
$$(2 + 6 + 12) + (10 + 18 + 12) = 20 + 40 = 60 \qquad (10)$$

The first sum, in Eq.(10), gives the number of amino acids, in this case, 2 sextets, 6 quartets and 12 doublets, as mentioned above. The second sum gives the respective degeneracies. Here, also, we could subtract Eq.(10) from the total number of codons to get

$$64 - (20 + 40) = 4 \qquad (11)$$

which is the number of stops, in this case four. In the following, we shall apply the *q*-deformations to the numbers in Eq.(8) and Eq.(10) to describe our classifications. Finally, let us mention that similar formulas, as Eqs.(6) and (10), albeit more complicated, could also be established for the various variants of the genetic code in Table 1 (Négadi, 2014).

## 4. CLASSIFICATIONS USING *q*-DEFORMATIONS

In this section, we use the *q*-deformed numbers of section 2 to establish several classication shemes, including not only of the non-standard (variant) genetic codes together with the standard one, but also the two known additional amino acids Selenocysteine and Pyrrolysine as well as a *prediction* of a minimal number of amino acids.

**4.1 The standard genetic code as a starting point**

Let us consider Eq.(8) for the standard genetic code, the total number of amino acids and their total degeneracy, separately. *Firstly*, we *q*-deform the total degeneracies, only, *keeping the number of amino acids, 20, unchanged*. We have

$$20 + \left([15]_q + [15]_q + [9]_q + [2]_q\right) \tag{12}$$

Using, in a first step, the defining relations (I), or Eq.(1) for the *q*-numbers (assuming here $q \geq 0$), we get

$$20 + [4 + 4q + 3(q^2 + q^3 + q^4 + q^5 + q^6 + q^7 + q^8) \\ + 2(q^9 + q^{10} + q^{11} + q^{12} + q^{13} + q^{14})] \tag{13}$$

It is clear, from Eq.(13), that the standard genetic code, our starting point, corresponds to $q=1$ (undeformed). In this case, we have 20+41=61, see Eq.(8). Now, one could verify that the following *q*-values describe all the variant genetic codes in the four categories $C_i$ of Table 1:

- $C_4$: $q \sim 0.9959$, *The Vertebrate Mitochondrial Code, The Thraustochytrium Mitochondrial Code* (Eq.(13)~60)[1]
- $C_3$: $q=1$, *The Standard Code, The Bacterial, Archeal and plant Plastide Code, The Alternative Yeast Nuclear Code, The Scenedesmus obliqus Mitochondrial Code* (Eq.(13)~61)

---

[1] In reference (Négadi, 2015), The *Thraustochytrium Mitochondrial code*, with four stop codons, was mistakenly placed in category $C_3$.

- **C$_2$**: $q\sim1.0040$, *The Yeast Mitochondrial Code, The Mold, Protozoan, and Coelenterate Code and the Mycoplasma/Spiroplasma Code, The Invertebrate Mitochondrial Code, The Echinoderm and Flatworm Mitochondrial Code, The Euploid Nuclear Code, The Ascidian Mitochondrial Code, The Chlorophycean Mitochondrial Code, The Trematode Mitochondrial Code, The Pterobranchia Mitochondrila Code, The Candidate Division SR1 and Gracilibacteria Code* (Eq.(13)~62)
- **C$_1$**: $q\sim1.0079$, *The Ciliate, Dasycladacean and Hexamita Nuclear Code, The Alternative Mitochondrial Code* (Eq.(13)~63)

The four $q$-values above lead to total degeneracy numbers very close to integer values in the second (deformed) part of Eq.(13): ~40 for C$_4$, ~42 for C$_2$ and ~43 for C$_1$. Note that we would have considered, instead, the following *fine-tuned* values for these $q$-values above **C$_4$**: 0.995884040159144, **C$_2$**:1.00398397186196, **C$_1$**:1.00784388952648, in which case we obtain integer values to very high precision. Summarizing and considering the corresponding (rearranged) equations like (9 and (11), we have

$$\mathbf{C_4} \ q\sim0.9959: \quad 20 + 40 + 4 = 64 \qquad (14.1)$$

$$\mathbf{C_3} \ q = 1: \qquad 20 + 41 + 3 = 64 \qquad (14.2)$$

$$\mathbf{C_2} \ q\sim1.0040: \quad 20 + 42 + 2 = 64 \qquad (14.3)$$

$$\mathbf{C_1} \ q\sim1.0079: \quad 20 + 43 + 1 = 64 \qquad (14.4)$$

In the above relations, the first number, 20, is the number of amino acids. The second one corresponds to the number of degenerate codons and finally the third one gives the number of stops. All the variant genetic codes are therefore classified, in one equation. *Secondly*, we deform the number of amino acids, *keeping the total number of degenerate codons, 41, unchanged*. Sill using equation (8), we have

$$\left([3]_q + [5]_q + [9]_q + [1]_q + [2]_q\right) + 41 \qquad (15)$$

Using the defining relations (I), or Eq.(1), we get

$$[5 + 4q + 3q^2 + 2q^3 + 2q^4 + q^5 + q^6 + q^7 + q^8] + 41 \qquad (16)$$

Of course, for $q=1$, we recover Eq.(8) for the standard genetic code: 20+41=61. Deforming slightly Eq.(16) we could get a *greater* number of amino acids. As a matter of fact, the following *q*-values *q*~1.0193 and *q*~1.0373 give a number of amino acids, corresponding to the above deformed expression in brackets, of 21 and 22, respectively. (more accurate values of *q* would be 1.01927498511164 and 1.03722675461874, respectively, to get with high precision the integer values 21 and 22, as the number of amino acids are discrete quantities.) These last two cases could describe nicely the inclusion of the experimentally known *additional amino acids* Selenocysteine (Sec), the 21$^{st}$ amino acid, *or/and* Pyrrolysine (Pyl), the 22$^{nd}$ amino acid. These are rare amino acids which have been discovered respectively in 1986 (see Böck et al., 1991) and in 2002 (Srinivasan et al., 2002). Both are cotranslationally encoded respectively by the codons UGA and UAG, which usually function as stop codons in the standard genetic code. If only one of them is considered, *at a time*, then the first *q*-value (*q*~1.0193) could describe it and in this case the number of amino acids is 21 (20 canonical+1 additional, Sec or Pyl). Also, as they are both encoded by (usually) stop codons, the total degeneracy in this case is unchanged and we have 21+41=62 coding codons and 2 remaining stop codons. Now, it has been shown (Zhang and Gladyshev, 2007) that these two additional amino acids could be *both* coded, in symbiotic delta proteobacteria of a gutless worm, *Olavius Algarvensis*. In this case, we have 22 amino acids, 20 canonical and 2 additional Sec *and* Pyl, and the value *q*~1.0373 is appropriate for the description of two additional amino acids. There are therefore 63 (22+41) coding codons and 1 stop codon left, in this case. *Thirdly*, we consider the minimal value of the deformation parameter *q* which, according to our assumptions (*q*≥0), is equal to 0. In this case, from Eq.(16), we have that the minimal number of amino acids is equal to 5. This (minimal) value agrees with several old and more recent claims about the smallness of the "primordial" amino acids number involved, at the origin of life, see for example, Woese (1965), Wong (1965), Crick (1968), Di Giulio (2008), Osawa (1992), Riddle et. al. (1997). Let us mention also that for values of *q* between 0 and 1, all the amino acids numbers between 5 (q=0) and 20 (q=1) could be modeled. We have that 6 amino acids correspond to *q*~0.2109, 7 amino acids to *q*~0.3647 and so on till 19 amino acids corresponding to *q*~0.9793. We could therefore conceive that, at the origin, *i.e.*, at *q*=0, the genetic code started with 5 amino acids and 59 degenerate-and-stop codons and, progressively as q varied *q*=0→*q*~0.2109→*q*~0.3647→*q*~0.9793→…→*q*=1, it reached its present ("canonical") form with 20 amino acids and 44 degenerate-and-stop codons, and more if we include the additional amino acids Selenocysteine and Pyrrolysine (*q*~1.0193 and *q*~1.0373, see above).

Now, we turn to the use of the second kind of *q*-numbers, in Eq.(II), or Eq.(3), and do the same calculations as above. First, we consider the deformation of the total degeneracy, *keeping the number of amino acids, 20, unchanged*. We have (the equivalent of Eq.(13))

$$20 + [3 + q + q^{-1} + 3(q^2 + q^{-2} + q^4 + q^{-4} + q^6 + q^{-6} + q^8 + q^{-8}) \qquad (17)$$
$$+ 2(q^{10} + q^{-10} + q^{12} + q^{-12} + q^{14} + q^{-14})]$$

Of course, here also, $q=1$ corresponds to the usual standard genetic code 20+41=61. However, for $q \neq 1$, there is a problem. As a matter of fact, the two values $q \sim 1.0287$ and $q \sim 1.0407$ lead to 42 and 43 for the total degeneracy number, suited to describe the categories $C_2$ and $C_1$, but there exist no $q$-value for the category $C_4$ (*The Vertebrate Mitochondrial Code* and *The Thraustochytrium Mitochondrial Code*). This is due to the symmetry $q \to q^{-1}$ of the $q$-numbers of the second kind. For example, the $q$-value taken above, $q \sim 1.0407$, and its inverse, $q^{-1} \sim 0.9609$, lead to the same value, 43, for the total degeneracy number. Let us examine now the (deformed) amino acid number, as in Eq.(16). We have, in this case

$$[4 + (q + q^{-1}) + 3(q^2 + q^{-2}) + 2(q^4 + q^{-4}) + (q^8 + q^6 + q^{-6} + q^{-8})] + 41 \qquad (18)$$

For $q=1$, the standard genetic code case is recovered. Also, the values $q \sim 1.0856$ and $q \sim 1.1217$ give 21 and 22 for the number of amino acids and they could be suited to describe the additional amino acids Selenocysteine and Pyrrolysine, as above when using kind-I $q$-numbers. However, here also, there is a problem: there exist no (real) $q$-value giving a minimal amino acid number. (It could be shown that the imaginary value $q=i$ gives 2 for a minimum, but this is outside our assumption q real and $\geq 0$). A way out these problems when using kind-II $q$-numbers, *i.e.*, the impossibility of description of the two variant genetic codes in the first category $C_4$ and also the non-existence of a minimal number of amino acids (see above), is to switch to the Vertebrate Mitochondrial genetic code, the most important representative of the category $C_4$, *as a starting point*. This is examined in the following sub-section.

### 4.2 The Vertebrate Mitochondrial genetic code as a starting point

We start therefore from Eq.(10), giving the number of amino acids (20) and the total degeneracy (40) of the Vertebrate Mitochondrial genetic code, and use, in a first calculation in this sub-section, the $q$-deformed numbers of kind-I. As above, we deform, first, the total degeneracy number, *keeping the amino acid number unchanged*. We have

$$20 + \left([10]_q + [18]_q + [12]_q\right) \tag{19}$$

In the detail, we get

$$20 + [3(1 + q + q^2 + q^3 + q^4 + q^5 + q^6 + q^7 + q^8 + q^9) + 2(q^{10} + q^{11}) \\ + q^{12} + q^{13} + q^{14} + q^{15} + q^{16} + q^{17}] \tag{20}$$

As usual, for *q*=1, we recover the Vertebrate Mitochondrial genetic code total degeneracy number 40. The total number of coding codons is equal to 60 (20+40) and there are 4 stop codons (64-60). Note that this case describes also the second genetic code variant, the *Thraustochytrium Mitochondrial Code*, with the same numbers so that the category **C₄** corresponds to *q*=1. Now, for *q*≠1, we could describe the other categories. The values *q*~1.0038, *q*~1.0074 and *q*~1.0109 lead to the total degeneracy numbers 41, 42 and 43 for the categories **C₃**, **C₂** and **C₁**, respectively. Note that, here, the standard genetic code, in category **C₃**, appears only as a deformation of the Vertebrate Mitochondrial code (see the conclusion section). Now, we deform the amino acids number, *keeping the total degeneracy number unchanged*. We have

$$[3 + 3q + 2(q^2 + q^3 + q^4 + q^5) + q^6 + q^7 + q^8 + q^9 + q^{10} + q^{11}] + 40 \tag{21}$$

Of course, for q=1, we recover the number 20 for the amino acids (the deformed part). Here also, we could could describe the two additional amino acids Selenocysteine or/and Pyrrolysine (see above) with the following values *q*~1.0118 and *q*~1.0228. Finally, from Eq.(21), we have that the minimal value *q*=0 gives a minimal value for the amino acids number equal to 3. These results are similar to those obtained above, starting from the standard genetic code and using the same kind-I *q*-numbers, except that, here, we started from the Vertebrate Mitochondrial genetic code. Still in this latter situation and to end this sub-section, we use, now, the *q*-numbers of kind-II. As we have done above, we, first, deform the total degeneracy number, *keeping the number of amino acids, 20, unchanged*. We have from Eqs.(3) and (10)

$$20 + [3(q + q^{-1} + q^3 + q^{-3} + q^5 + q^{-5} + q^7 + q^{-7} + q^9 + q^{-9}) \\ + 2(q^{11} + q^{-11}) \\ + (q^{13} + q^{15} + q^{17} + q^{-13} + q^{-15} + q^{-17})] \tag{22}$$

From Eq.(22) we have that the values $q\sim1.0268$, $q\sim1.0379$ and $q\sim1.0464$ give 41, 42 and 43 for the total degeneracy number in the categories $C_3$, $C_2$ and $C_1$, respectively. Here, also, the standard genetic code, in the category $C_3$, appears only as a deformation of the Vertebrate Mitochondrial genetic code. Deforming, now, the amino acids number, *keeping the total degeneracy number unchanged*, we have from Eq.(10)

$$[3(q + q^{-1}) + 2(q^3 + q^5 + q^{-3} + q^{-5}) + q^7 + q^9 + q^{11} + q^{-7} + q^{-9} + q^{-11}] \\ + 40 \tag{23}$$

As usual, for $q=1$, we recover the number of the amino acids 20 and 20+40=60, but this last case appears to have much less description capability, compared to the other cases considered above. As a matter of fact, we could only describe the additional amino acids (Selenocysteine or/and Pyrrolysine) with the values $q\sim1.0568$ and $q\sim1.0804$, and there exist no (real) $q$-value leading to a minimal value for the amino acids number (even the imaginary value $q=i$ gives zero for this latter, see at the end of sub-section 4.1).

## CONCLUSION

In summary of this study, the two (*semi-phenomenological*) models based on the definition (I) of the non-symmetric $q$-numbers, either with the standard genetic code (case A of sub-section 4.1), or the Vertebrate Mitochondrial Genetic Code (case B of sub-section 4.2), as a starting point, seem better at describing (fitting) the total degeneracy numbers of the standard genetic code and its variants, the number of the additional 21[th] or/and 22[th] amino acids as well as at "predicting" a minimal number of amino acids (respectively 5 and 3, see above) suitable for describing a primordial life protein alphabet, than the models based on the symmetric definition (II) of $q$-numbers. The models using the symmetric q-numbers of kind II lack this last (predictive) capability. The above described models, possessing a "predictive" property, have been termed *semi-phenomenological* because, besides the fact that they can fit the experimental data (total degeneracies and additional amino acids), they also "predict" a minimal number of amino acids. Now, following the widely admitted idea that the non-standard genetic codes, including the Vertebrate Mitochondrial Genetic Code, are all slight variations of the standard (quasi-universal) genetic code, then we can, at last, exclude the case B and retain only the case A as our final semi-phenomenological classification model. As a closing remark, let us mention that we have recently built a similar classification model relying on the deformation of Fibonacci numbers (see Négadi, 2015).